\documentclass[12pt]{article}
\usepackage{amsmath,amssymb,epsfig}

\newcommand{\cl}{{\rm cl}}
\newcommand{\qu}{{\rm qu}}
\newcommand{\h}{{\rm h}}
\newcommand{\Z}{{\bf Z}}
\newcommand{\C}{{\bf C}}
\renewcommand{\thefootnote}{\fnsymbol{footnote}}

\begin{document}

\title{
\begin{flushright}
\ \\*[-80pt]
\begin{minipage}{0.2\linewidth}
\normalsize
%hep-th/yymmnnn \\
KUNS-2112 \\*[50pt]
\end{minipage}
\end{flushright}
{\Large \bf Higher Order Couplings \\ from
Heterotic Orbifold Theory
\\*[20pt]}}

\author{Kang-Sin Choi$^{1,}$\footnote{
E-mail address: kschoi@th.physik.uni-bonn.de} \ and \
Tatsuo~Kobayashi$^{2,}$\footnote{
E-mail address: kobayash@gauge.scphys.kyoto-u.ac.jp} \\*[20pt]
$^1${\it \normalsize
Physikalisches Institut, Universit\"at Bonn,
Nussalle 12, D-53115 Bonn, Germany} \\
$^2${\it \normalsize
Department of Physics, Kyoto University,
Kyoto 606-8502, Japan} }

\date{}

\begin{titlepage}
\maketitle
\thispagestyle{empty}
\begin{abstract}
We calculate couplings of arbitrary order from correlation
functions among twisted strings, using conformal field theory.
Twisted strings arise in heterotic string compactified on
orbifolds yielding matter fields in the low energy
limit. We calculate completely the classical and the
quantum amplitude including normalization, up to a contribution from
K\"ahler potential.
The classical action has saddle points which are interpreted as
worldsheet instantons described by metastable untwisted strings,
formed by twisted strings distributed at certain fixed points.
This understanding generalizes the area
rule, in the case that the locations of twisted strings do not form 
a polygon, and provides a general rule for calculating
these kinds of instanton corrections. An interpretation of couplings
involving linearly combined states is given, which commonly appear in
non-prime order orbifolds.
The quantum part of the amplitude is given by ratios of
gamma functions with order one arguments.

\end{abstract}
\end{titlepage}

\renewcommand{\thefootnote}{\arabic{footnote}}
\setcounter{footnote}{0}

\section{Introduction}

Superstring theory is a promising candidate for unified theory
including gravity. Heterotic orbifold construction is one of
interesting constructions for four dimensional string models \cite{Dixon,IMNQ}.
(See also for resent works
Ref.~\cite{Kobayashi:2004ud,Forste:2004ie} and for review
\cite{Choi-Km}.) One can solve equation of motion of string on the
orbifold background, and geometrical picture is clear in heterotic
orbifold models. Thus, several aspects can be computed and can be
understood from the geometrical viewpoint.

Heterotic orbifold models have modes localized at fixed points,
that is, twisted strings. 3-point couplings as well as 4-point
couplings of these localized modes have been computed analytically
\cite{Hamidi:1986vh,Dixon:1986qv,Atick:1987kd,Burwick:1990tu,Kobayashi:2003vi},
and the size of Yukawa coupling $Y$ is obtained as $Y\sim e^{-A}$, where $A$
denotes naively the area of triangle corresponding to three fixed points
of twisted strings. This will be clarified more and generalized in
this paper. This aspect is quite interesting from the
phenomenological viewpoint. One can obtain suppressed Yukawa
couplings when twisted strings are localized far away from each
other. That is, one could explain the hierarchy of quark and lepton
masses as well 
as their mixing angles when they are localized at different
places.

We have to study selection rules for allowed couplings
in order to examine whether realistic fermion masses and
mixing angels can be realized from string theory.
The space group selection rule \cite{Dixon:1986qv,Kobayashi:1991rp}
constrain allowed Yukawa couplings rather strongly.
For example, on prime order orbifolds off-diagonal Yukawa
couplings are not allowed, and we can not obtain realistic
mixing angles by using only 3-point couplings with the minimum number
of Higgs fields.\footnote{In non-factorizable orbifold models, 
off-diagonal Yukawa couplings are allowed, but it is still difficult 
to derive realistic Yukawa matrices \cite{Forste:2006wq}.}
On non-prime order orbifolds, off-diagonal Yukawa couplings are
allowed \cite{Kobayashi:1991rp} and possibilities
for leading to realistic quark and lepton masses and
mixing angles have been studied  \cite{Ko:2004ic}.
However, to realize fermion masses and mixing angles
in string theory is still a challenging issue.

In this paper, we study generic higher order couplings than
renormalizable couplings in heterotic orbifold models. 
Higher dimensional operators become effective Yukawa couplings
after symmetry breaking. Suppose that we have a coupling of type
$FfH\prod_i\phi_i$ in the superpotential of effective field theory, 
where $F$ and $f$ are chiral matter fields corresponding 
quarks and leptons, $H$ denotes electroweak Higgs superfields and 
$\phi_i$ correspond to several heavy modes.
When all scalar components of the superfields $\phi_i$ 
develop their vacuum expectation values (VEVs), 
this higher dimensional operator
becomes a Yukawa coupling among chiral fermions $F$ and $f$ and
the electroweak Higgs fields $H$.
Thus, there is a possibility for deriving
quark/lepton masses and mixing angles through this type
of symmetry breaking, but by use of not only 3-point couplings.
Indeed, such possibility has been examined in explicit models
\cite{IMNQ,Kobayashi:2004ud,Forste:2004ie}.
Therefore, it is important to study
selection rules of allowed higher order couplings and
compute magnitude of allowed couplings.\footnote{
It would also be useful to study non-Abelian flavor symmetries
\cite{Kobayashi:2006wq,Ko:2007dz} and accidental global symmetries
\cite{Choi:2006qj} in string models.}
When $\phi_i$ correspond to localized modes on orbifold 
fixed points, the above effective Yukawa coupling may 
correspond to a 3-point coupling on a Calabi-Yau manifold, 
where orbifold singularities are smoothed by the VEVs 
of $\phi_i$.
Thus, calculations of higher order couplings on 
the orbifold are also important from the viewpoint of calculations 
of 3-point couplings on the Calabi-Yau manifold 
around the orbifold limit.

We compute magnitudes of $L$-point couplings.
Generic aspects of $L$-point couplings heterotic orbifold models
have been obtained in \cite{Atick:1987kd}.
Here we apply it to concrete heterotic orbifold models.
Similar calculation has been carried out for
generic $L$-point couplings in intersecting D-brane
models \cite{Abel:2003yx}.\footnote{See for 3-point couplings in
intersecting D-brane models
\cite{Cremades:2003qj,Cvetic:2003ch,Abel:2003vv,{Klebanov:2003my}}.}

The paper is organized as follows.
In section 2, we give a brief review on heterotic orbifold models
in order to fix our notation.
Then, we study the selection rule due to
discrete $R$-symmetry and the space group.
In section 3, we compute classical contributions of $L$-point couplings.
Their quantum parts are calculated in section 4.
In section 5, we consider normalization of correlation functions.
Section 6 is devoted to conclusion and discussion.
In appendix, we give useful formulae for hypergeometric functions and
their multivariable generalizations.

\section{Setup}

\subsection{Twisted strings and their vertex operators}

The heterotic string theory consists of 10D right-moving superstring
and 26D left-moving bosonic string. For the common ten (bosonic)
dimensions, we consider the background with our 4D space-time and 6D
orbifold. The other 16D left-moving bosonic string correspond to a
gauge part. A 6D orbifold is a division of a 6D torus $T^6$ by a
twist $\theta$, while $T^6$ is obtained as ${\bf R}^6/\Lambda$, where
$\Lambda$ is a 6D lattice.
The twist $\theta$ must be an
automorphism of the lattice $\Lambda$, and its eigenvalues are
diag$(e^{2 \pi i \eta_1}, e^{2 \pi i \eta_2},e^{2 \pi i \eta_3})$ in the
complex basis $Z_i$ $(i=1,2,3)$.
We mainly concentrate ourselves to the case that $T^6$ is
factorizable as $T^2\times T^2 \times T^2$.
To preserve 4D $N=1$ supersymmetry
(SUSY), they must satisfy the following condition,
\begin{equation}
\eta_1+\eta_2+\eta_3={\rm~integer},
\end{equation}
where $\eta_i$ is not integer for each $i=1,2,3$.

The twisted string is a closed string up to orbifold identification
\begin{equation} \label{twisted}
 Z(e^{2\pi i}z,e^{-2\pi i}\bar z) = \theta^k Z(z,\bar z) + v, \quad
 v \in \Lambda,
\end{equation}
where $\Lambda$ is the above lattice (in the complex basis) 
defining the orbifold. It makes
sense to restrict the phase to be $-1 \le k \eta_i \le 1$. Its zero
mode satisfies the same condition, and it is called a fixed point on
the orbifold. The fixed point can be represented by the
corresponding space group element, $(\theta^k,v)$. Note that the
fixed point $(\theta^k,v)$ is equivalent to
$(\theta^k,v+(1-\theta^k)\Lambda)$. They belong to the same
conjugacy class. The sector with $k=0$ corresponds to the so-called
untwisted sector.

The local operator called the twist operator $\sigma_k(z)$ takes
into account the nontrivial boundary condition (\ref{twisted}) by
inducing a branch point at $z$ with the order $k/N$ on the
world-sheet, but the theory remains local by moding out by
orbifold projection. The ground state corresponding to the twisted
string on the fixed point $(\theta^k,v)$ is generated from the
untwisted ground state $|0\rangle$ by the twist field
$|\sigma_k\rangle = \sigma_k(0, 0) |0 \rangle$. These twist fields
have the operator product expansions (OPEs)
\begin{align}
 \partial Z(z) \sigma_{k}(0, 0) &\sim z^{k/N-1}\tau_{k}(0,0),
 \label{OPEdZsigma1} \\
 \bar \partial Z(\bar z) \sigma_{k}(0,0) &\sim \bar z^{-k/N}\tilde
\tau_{k}'(0,0), \label{OPEdZsigma2}
\end{align}
which are understood as the most singular parts in the mode
expansion for $0 \le \frac{k}{N} \le 1$. For the other region $-1
\le \frac{k}{N} \le 0$, we have the corresponding relations by
replacing $k$ with $N-k$. Also we have similar expressions for
$\overline Z$. Their conformal weights for the holomorphic and
anti-holomorphic parts are
\begin{equation} \label{confwght}
h_{\sigma_k} = h_{\sigma_{N-k}} = \frac12 {k \over
  N} \left(1-{k \over N} \right) ,
\end{equation}
thus inducing a shift of zero point energy.

Each $\theta^k$-twisted sector has several ground states, that is,
twist fields corresponding to several fixed points under $\theta^k$
twist.
When we specify the fixed point $f$, we denote
$\sigma_{f,k}$.
Also we use the notation $\sigma_{(\theta^k,v)}$, where
$(\theta^k,v)$ denotes the space group element
corresponding to the fixed point $f$ under $\theta^k$ twist.

On non-prime order orbifolds, fixed points under higher twist
$\theta^k$ ($k > 1$)
are not always fixed under $\theta$ or
twist fields $\sigma_{(\theta^k,v)}$ are not always
eigenstates of the twist $\theta$.
To make eigenstates, we have to take the following
linear combinations \cite{Kobayashi:1990mc,Kobayashi:1991rp},
\begin{equation} \label{lincombi}
\sigma_{(\theta^k,v)}^{(\gamma)} \equiv \frac{1}{\sqrt{k}} \left(
\sigma_{(\theta^k,v)}
+\gamma \sigma_{(\theta^k,\theta v)} + \gamma^2 \sigma_{(\theta^k,\theta^2 v)}
+ \cdots \gamma^{k-1} \sigma_{(\theta^k,\theta^{k-1} v)} \right),
\end{equation}
where $\gamma =e^{2\pi i\ell /m}$ with integer $\ell$ to be determined
by gauge quantum numbers and internal momenta.
This linear combination may include twist fields corresponding 
to fixed points, which belong to the same conjugacy class.

We consider the covariant quantization with the explicit conformal and
superconformal ghosts.
It is convenient to bosonize right-moving fermionic string and write
bosonized degrees of freedom by $H^t(\bar z)$. In the bosonized formulation,
untwisted massless modes have momenta $p_t$ for $t=1,\cdots,5$,
which are quantized on the $SO(10)$ weight lattice. The space-time
boson and fermion correspond to $SO(10)$ vector and spinor,
respectively. The compact space corresponds to $SO(6)$. The twisted
sector $T_k$ has shifted $SO(6)$ momenta, $r_i=p_i+k\eta_i$, which are
often called $H$-momenta.

A bosonic massless state has the corresponding vertex operator,
\begin{equation}
V_{-1} = e^{-\phi}\prod_{i=1}^3
(\partial^{m_i} Z_i)^{{\cal N}_i} (\partial^{\bar m_i} \bar
Z_i)^{\bar {\cal N}_i}e^{ir_tH_t}e^{iP^IX^I}e^{ikX}
\sigma_{(\theta^k,v)}^{(\gamma)} ,
\end{equation}
naturally in the $(-1)$-picture, where $\phi$ is the bosonized
ghost, $P^IX^I$ corresponds to the gauge part and $kX$ corresponds
to 4D part. Here, $\partial^{m_i} Z_i$ and $\partial^{\bar m_i} \bar
Z_i$ denote oscillators for the left-mover, and ${\cal N}_i$ and $\bar
{\cal N}_i$ are oscillator numbers, which these massless modes include.
Similarly, we can write massless modes
corresponding to space-time fermions as
\begin{equation}
V_{-\frac12} = e^{-\frac12 \phi} \prod_{i=1}^3
(\partial^{m_i} Z_i)^{{\cal N}_i} (\partial^{\bar m_i} \bar
Z_i)^{\bar {\cal N}_i}
e^{ir_t^{(f)}H_t}e^{iP^IX^I}e^{ikX}
\sigma_{(\theta^k,v)}^{(\gamma)},
\end{equation}
in the $(-\frac12)$-picture. We understand that the $H$ fields
contains the four dimensional spin field.
The $H$-momenta for space-time fermion and boson, $r_i^{(f)}$ and
$r_i$ in the same supersymmetric multiplet are related each other
\begin{equation}
r_i = r_i^{(f)} + (1,1,1,1,1)/2,
\end{equation}
that is, $(1,1,1,1,1)/2$ corresponds to the $H$-momentum of
unbroken 4D space-time SUSY charge.
To each vertex operator, we have to include overall normalization
\begin{equation} \label{vertexnorm}
 g_c \prod_{i=1}^3 \left[ \left( {2 \over \alpha'} \right)^{1/2}
{i \over (m_i-1)!} \right]^{{\cal N}_i}
 \left[ \left( {2 \over \alpha'} \right)^{1/2} 
{i \over (\bar m_i-1)!} \right]^{\bar {\cal N}_i} , 
\end{equation}
from the state-operator mapping or the unitarity relation \cite{Po}.
The closed string coupling
$g_c$ is expressed in terms of ten dimensional gauge and gravitational
couplings 
\begin{equation} \label{clstrcoupling}
 g_c = {\alpha^{\prime 1/2} g_{\rm YM} \over 4 \pi} = {\kappa
  \over 2 \pi}.
\end{equation}
Thus, including one more field suppresses the corresponding
coupling by one inverse mass dimension ${\cal O}(\alpha'^{1/2})$ as we expect.
We have omitted the two-cocycles, which determine the overall sign
\cite{Goddard:1986bp}.

We calculate a correlation function among $L$ twisted matter fields
including two space-time fermions
on the $\Z_N$ orbifold, along the
lines \cite{Dixon:1986qv,Atick:1987kd,Burwick:1990tu,Abel:2003yx}.
It yields a higher
order coupling in the zero momentum limit $k_i \to 0$. Since
the background has the superconformal ghost charge 2, the correlation
function is of the form,
\begin{equation}
\lim_{k_i \to 0} e^{-\lambda} \int \prod_{i=4}^L dz_i  \big \langle :c \tilde c V_{-1}(z_1,\bar z_1)::c \tilde c
V_{-\frac12}(z_2, \bar z_2):: c
\tilde c V_{-\frac12} (z_3, \bar z_3): \prod_{i=4}^{L} :V_{0}(z_i,\bar
z_i): \big \rangle, 
\end{equation}
such that the total ghost charge vanishes. Here $\lambda$ is
worldsheet cosmological constant and we also have three
bosonic ghost fields $\tilde c(\bar z)$ and $c(z)$. We take radial
ordering implicitly, which reflects the ordering property from 
noncommutative space group.

In order to make the total superconformal ghost charge vanishing, 
we need vertex
operators $V_0$ in the $0$-picture. We can obtain $V_0$ by operating the
following picture changing operator on $V_{-1}$
\cite{Friedan:1985ge},
\begin{equation}
Q=e^\phi (e^{-2 \pi i r^v_i \cdot H}\bar \partial Z_i + e^{2 \pi i r^v_i
\cdot H}\bar \partial \bar Z_i), \label{p-change}
\end{equation}
where $r^v_1=(1,0,0)$, $r^v_2=(0,1,0)$ and $r^v_3=(0,0,1)$ 
for the components corresponding to the 6D compact space. 
Thus we have
\begin{equation} \begin{split}
  V_0 =& (\alpha'/2)^{1/2}  e^{i r_t
  \cdot H_t} e^{i P^I X^I} e^{i k \cdot X} \\
 & \times \prod_{i=1}^3 
(\partial^{m_i} Z_i)^{{\cal N}_i} (\partial^{\bar m_i} \bar Z_i)^{\bar
  {\cal N}_i}  \left[i k \cdot \psi 
  \sigma^{\gamma}_{(\theta^k,v)} + \sum_{i=1}^3 \left( e^{-2 \pi i r^v_i\cdot H}
   \bar \partial Z_i  +  e^{2 \pi i r^v_i\cdot H}
   \bar \partial \bar Z_i \right)
\tau^{(i)}\prod_{j=1,j \ne i}^3 \sigma^{(j)} \right] ,
\end{split} \end{equation}
up to the same normalization (\ref{vertexnorm}). 
Here, $\sigma^{(j)}$ is the
$j$th component of $\sigma^\gamma_{(\theta^k,v)}$.
Containing no derivatives, the higher order coupling is defined in the
zero momentum limit $k \to 0$. 
Thus the first term is not relevant. The only change is that some
components of twist fields $\sigma$ are replaced by excited twist fields
$\tau$, and the normalization factors $( \alpha'/2)^{1/2}$, in accord
with the number of oscillators in (\ref{vertexnorm}). In the
next subsection, we see this change does not modify the calculation in
the case that all the twist fields are simply $\sigma$, not excited twist fields.

\subsection{Selection rules} \label{subsec:selrule}

Here we briefly summarize the selection
rules \cite{Dixon:1986qv,Kobayashi:1991rp,Kobayashi:2004ud,
Lebedev:2007hv,CKNRV}.
The vertex
operator consists of several parts, the 4D part $e^{ikX}$, the gauge
part $e^{iPX}$, the 6D twist field
$\sigma_{(\theta^k,v)}^{(\gamma)}$, the 6D left-moving oscillators
$\partial Z_i$ and the bosonized fermion $e^{irH}$, as explained in
the previous subsection. Each part has its own selection rule for
allowed couplings. The selection rules of the 4D part and the gauge
part are simple, that is, the 4D total momentum $\sum k$ and the
total momentum of the gauge part $\sum P$ should be conserved. The
latter rule is nothing but the requirement of gauge invariance. The
other parts lead to non-trivial selection rules. In this subsection,
we study the selection rule from the $H$-momenta and oscillators,
as well as  the selection rule from the 6D
twist fields $\sigma_{(\theta^k,v)}^{(\gamma)}$.

The total $H$-momentum should be conserved like the 4D momentum and
the gauge momentum $P$. For example, for 3-point couplings $\langle
V_{-1}V_{-1/2}V_{-1/2} \rangle$, they should satisfy the following
condition,
\begin{equation}
\sum r_i = 1.
\end{equation}
Here we take a summation over the $H$-momentum for the scalar
components, using the fact that the $H$-momentum of fermionic
component differs by $-1/2$.

Another important symmetry is the twist symmetry of
oscillators.
We consider the following twist of oscillators,
\begin{eqnarray} \label{oscitransf}
& & \partial Z_i \rightarrow e^{2 \pi i \eta_i}\partial Z_i, \qquad
\partial \bar Z_i \rightarrow e^{-2 \pi i \eta_i}\partial \bar Z_i, \\
& & \bar \partial Z_i \rightarrow e^{2 \pi i \eta_i}\bar \partial Z_i,
\qquad \bar \partial \bar Z_i \rightarrow e^{-2 \pi i \eta_i}\bar
\partial \bar Z_i,
\end{eqnarray}
without summation over each $i=1,2,3$.
Allowed 3-point couplings $\langle V_{-1}V_{-1/2}V_{-1/2} \rangle$
should be invariant under the above $\Z_N$ twist.

However, for generic $L$-point couplings we have to carry out picture
changing, and the picture changing operator $Q$ includes
non-vanishing $H$-momenta and right-moving oscillators $\bar
\partial Z_I$ and $\bar \partial \bar Z_i$. 
Thus, the definition of $H$-momentum depends on the choice of the 
picture.
However, the R-charges, which are defined
as \cite{Kobayashi:2004ud}\footnote{See
also \cite{Araki:2007ss} and references therein.}
\begin{equation}
R_i \equiv r_i + {\cal N}_i - \bar {\cal N}_i,
\end{equation}
are invariant under picture-changing. Here we do not
distinguish oscillator numbers for the left-movers and right-movers,
because they have the same phase under $\Z_N$ twist. Indeed,
physical states with $-1$ picture have vanishing oscillator number
for the right-movers, while the oscillator number for the
left-movers can be non-vanishing. Thus, hereafter   ${\cal N}_i$ and
$\bar {\cal N}_i$ denote the oscillator number for the left-movers,
because we study the physical states with $-1$ picture from now. For
simplicity, we use the notation $\Delta {\cal N}_i = {\cal N}_i -
\bar {\cal N}_i$. Now, the selection rule due to R-symmetry is
written as
\begin{equation}
\sum R_i = 1 \quad {\rm mod} \quad N_i,
\end{equation}
where $N_i$ is the minimum integer satisfying $N_i = 1/\hat \eta_i$,
where $\hat \eta_i= \eta_i + m$ with any integer $m$. For
example, for $\Z_6$-II orbifold, we have $\eta_i=(1,2,-3)/6$, and
$N_i=(6,3,2)$.

Whereas the twist operator $\sigma_{k}$ itself does not transform
under the twist of oscillators,
the excited twist operator $\tau_k$ transforms like
the oscillator in (\ref{oscitransf}), since it is nothing but the
product of an oscillator and a twist operator, from the
transformational point of view. The modified $H$-momentum has
a compensating property and the resulting amplitude is invariant
under the twist of oscillators,
as it must be because the picture changing operator
is invariant. However the OPE $\partial Z \tau_k$ is of a similar form
of a twisted operator
\begin{equation} \label{OPEXtau} \begin{split}
 \partial Z(z) \tau_{k}(0, 0) &\sim z^{k/N-1}\upsilon_{k}(0,0), \\
 \bar \partial Z(\bar z) \tau_{k}(0,0) &\sim \bar z^{-k/N}\tilde
\upsilon_{k}'(0,0),
\end{split} \end{equation}
which is readily extracted from the OPE $\partial Z \partial Z
\sigma_k$. It has a branch structure like $z^{2k/N-2}$, but the part
$z^{k/N-1}$ is carried by $\tau_k$ from its definition to leave
(\ref{OPEXtau}). Thus the amplitude including excited twisted
operators is the same as one including only twisted operators, up
to the overall normalization.

We have the space group selection rule. Here, we study the selection
rule for twist fields, $\sigma_{(\theta^k,v)}^{(\gamma)}$. First of
all, the product of $\Z_N$ phases $\gamma$ should satisfy $\prod
\gamma =1 $.\footnote{That is automatic for physical states when 
the other selection rules are satisfied \cite{Lebedev:2007hv,CKNRV}.}
Next, we study the space group selection rule. Now, let us consider
$L$-point couplings of twisted states corresponding to
$(\theta^{k_{i}},v_{i})$ ($i=1,\cdots, L$). Their couplings are
allowed if the product of space group is the identity, i.e.,
\begin{equation}
\prod_{i=1}^L (\theta^{k_{i}},v_{i}) = (1,0).
\end{equation}
{ Since space group elements do not commute, nor do vertex
operators, the ordering of vertex operators in
the coupling is important.} We have to take into
account the fact that $(\theta^k,v)$ is equivalent to
$(\theta^k,v+(1-\theta^k)\Lambda)$. Thus, the condition for allowed
couplings is that the product of space group elements must be the
identity up to such equivalence. The space group selection rule
includes the point group selection rule, which requires $\prod
\theta^{k_{i}} =1$, i.e. $\sum k_{i}= 0$ (mod $N$) for the $\Z_N$
orbifold. 
The rules for the linearly combined states are discussed in
detail in Ref \cite{CKNRV}.

\section{The classical contribution}

Here we consider the 6D $\Z_N$ orbifolds, which can be factorized
as three 2D $\Z_N$ orbifolds,
and we concentrate ourselves to calculation of correlation
functions on the 2D $\Z_N$ orbifolds.
The following analysis can be extended to other cases,
where 6D $\Z_N$ orbifold is not factorizable or
6D $\Z_N$ orbifold includes 4D non-factorizable orbifold.

The nontrivial part is the correlation function among $L$ twist operators
\begin{equation} \label{corrfn}
 {\cal Z} \equiv \langle \sigma_{k_1} \sigma_{k_2} \dots \sigma_{k_L}
 \rangle.
\end{equation}
This can be calculated independently of the remaining components of
vertex operators, because they commute.
Some of them should be excited twisted states $\tau_k$ for
the total ghost charge being $-2$. Indeed we noted above that the
amplitude is the same with a number of factors $(2 \alpha^\prime)^{-1/2}$.

From the point group selection rule we have
\begin{equation}
 \sum_{i=1}^L {k_{i} \over N} \equiv M,
\end{equation}
where $M$ must be an integer. For the moment we {\em assume}
\begin{equation}  \label{sumk}
 M = L-2,
\end{equation}
and we will relax this condition later. This choice is the most
convenient one because of two reasons.
First, by doubling trick we can relate the
corresponding amplitude with that of open strings \cite{Abel:2003yx},
where this is the closedness condition for $L$
sided polygon. This gives rise to the generalized
Schwarz--Christoffel transformation and the area rule that we will show below.
Another reason is that we can obtain the closed form of integration,
in terms of a multivariable hypergeometric function \cite{Opdam}.

If we use path integral formulation, the correlation function
(\ref{corrfn}) is divided as  
\begin{equation} \label{corrclandqu}
  {\cal Z} = {\cal Z}_{\qu} \cdot \sum_{\{Z_{\cl}\}} \exp(-S_{\cl}),
\end{equation}
according the classical value and the quantum fluctuation around it,
i.e. $Z = Z_{\cl} + Z_{\qu}$. 

The corresponding tree-level Feynman diagram is a sphere with a
number of vertex operators inserted. Every object in string theory,
including the vertex operators and the correlation function
(\ref{corrfn}) can be separated into holomorphic and antiholomorphic
part. Considering one of them, holomorphicity restricts many things
in a very simple form. Using the compactification $\C
\cup \{ \infty\} = S^2$ we can cover all the coordinate on the
sphere by holomorphic coordinate $z$ except infinity. To take care
of infinity we introduce another patch $z = 1/u$ where $z \to \infty$
becomes $u=0$. Considering a holomorphic solution from
\begin{equation} \label{dropsmin2}
 \partial_u Z = - z^2 \partial Z
\end{equation}
if we need the LHS well-behaved at $u=0$, in the RHS $\partial Z$
should drop faster than $z^{-2}$ as $z$ goes to infinity.

\subsection{$L$-point coupling}

We begin calculating the classical
contribution first. We see
that the classical solution is the completely factorized
part for each inserted operator.
Therefore, from the OPEs (\ref{OPEdZsigma1}) and (\ref{OPEdZsigma2}),
the classical solutions are obtained 
by the holomorphicity and the desired singular structures as
\begin{equation} \label{clansatz}
 \begin{split}
   \partial Z_{\cl} (z) &= a \omega(z), \\
  \bar \partial Z_{\cl} (\bar z) & = \sum_{l=2}^{L-2} b_l \bar
  \omega^{\prime l}(\bar z).
 \end{split}
\end{equation}
Here we define the basis of  $(L-2)$ functions
\begin{equation} \label{omegas} \begin{split}
    \omega(z)  &= \prod_{i=1}^L (z-z_i)^{k_i/N-1}, \\
    \bar \omega^{\prime l}(\bar z) &= \prod_{i=1}^L (\bar z-\bar z_i)^{-k_i/N}
      \prod_{j=2,j\ne i}^{L-2} (\bar z-\bar z_j), \quad
    i=2,\dots,L-2.
\end{split} \end{equation}
For the symmetric orbifold they are complete since $\bar
\partial \overline Z_{\cl} (\bar z) = (\partial Z_{\cl} (z))^*$ and
$ \partial \overline Z_{\cl} (\bar z) = (\bar \partial Z_{\cl} (z))^*$.
From (\ref{dropsmin2}), the whole part (\ref{clansatz})
should behave as $z^{-2}$ as $z \to \infty$. Using (\ref{sumk}),
we can see $\omega(z)$ does. For $\bar \omega^{\prime l}(\bar z)$,
we admitted an additional degree of freedom, i.e. changing the power
of $z-z_j$ singularity by integer, since it does not modify the
branch cut structure. Hence, we have many free parameters $b_l$.

\begin{figure}[t]
\begin{center}
\includegraphics[height=3cm]{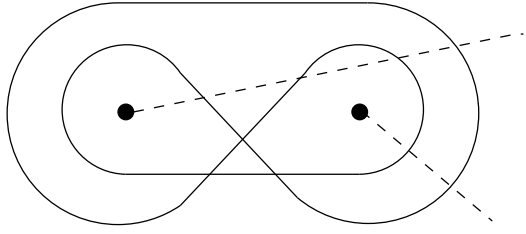}
\caption{Pochhammer loop. Whatever direction the branch cut we have,
  the nontrivial loop goes in and out respectively exactly once
  through each cut, enclosing two branch points.} \label{pochloop}
\end{center}
\end{figure}

To determine the coefficients $a,b_l$, we should consider the
global monodromy condition. The relation (\ref{twisted}) does not
take into account the global phase if we transport branch cuts
from more than one fields. Taking a contour $C$ encircling more
than one points gives the relation between overall coefficients in
(\ref{clansatz}) and net translation $v$ in the target space
\begin{equation} \label{glmonodef}
 \Delta_{C} Z = \oint_{C} dz \partial Z + \oint_{C} d\bar z \bar \partial Z
 = v.
\end{equation}
This expression makes sense only if there is no additional phase.
The quantum part does not carry any amplitude $\Delta_{C} Z_{\qu} =
0$.
Hence, Eq.~(\ref{glmonodef}) shows a purely classical contribution.
Upon integration relating these branch cuts,
the Pochhammer loop \cite{Poch} is a clever way to encompass the two
branch points nontrivially without phase. For each
branch cut the contour goes in and out exactly once through the
cuts, depicted in Fig. \ref{pochloop}. Its effect is to encircle
the fixed points: $f_1$ clockwise, $f_2$ counterclockwise, $f_1$
counterclockwise and then $f_2$ clockwise. 
In terms of space group elements
$(\omega,v'_1 = (1-\omega)(f_1+v_1))$ and $(\theta,v'_2 =
(1-\theta)(f_2+v_2))$ with $v_{1,2} \in \Lambda$, we obtain
\begin{equation} \label{pochtransl} \begin{split}
 &(\theta,v'_2)(\omega,v'_1)^{-1}(\theta,v'_2)^{-1}(\omega,v'_1) \\
 &=\left(1,(1-\omega^{-1})v_2 + (1-\theta)\omega^{-1} v_1 \right)\\
 &=\left(1,(1-\omega^{-1})(1-\theta)(f_2-f_1+v) \right),
\end{split} \end{equation}
where $v=v_2 -v_1$. The encircling is not necessarily once, i.e,
$\omega \ne \theta$ in general, in which we cannot draw branch
cuts. The net effect is pure translation. It turns out that every
contour is generated by the basis of Pochhammer loops $C_i$ encircling
$i$-th and $(i+1)$-th points.

Taking Pochhammer loops $C_i$ around the
vertices $z_i$ and $z_{i+1}$, from (\ref{pochtransl}) we obtain
\begin{equation} \begin{split}
  \Delta_{C_i} Z_{\cl} &= \oint_{C_i} dz \partial Z_{\cl} (z) +
  \oint_{C_i} d \bar z \bar \partial Z_{\cl} (\bar z) \\
   &= (1-e^{-2\pi i k_i/N}) (1-e^{2\pi i k_{i+1}/N}) (f_{i+1}-f_i+ v) \\
   &= 4 e^{-\pi i(k_i-k_{i+1})/N} \sin \left({ k_i \pi \over N} \right)
  \sin \left({ k_{i+1} \pi \over N} \right)(f_{i+1}-f_i+v).
\end{split}
\end{equation}
We have $(L-2)$ vectors $f_{i+1}-f_i +v$ and $L$ angles (with the
constraint (\ref{sumk})), which completely specify $L$-sided
polygon.

Later we can express the solution in terms of the following
integrals
\begin{equation} \label{wmatrix}
 W_{l}^1 \equiv \oint_{C_l} d z \omega (z), \quad
 W_{l}^i \equiv \oint_{C_l} d \bar z \bar \omega_i (\bar z),
\end{equation}
and
\begin{equation}  \label{glmonodromy}
 W_l^i = (1-e^{-2\pi i k_i/N}) (1-e^{2\pi i k_{i+1}/N}) F^i_l,
\quad i=1,\dots,L-2,
\end{equation}
with
\begin{equation} \label{Fmatrix}
    F_l^1 = \int_{z_l}^{z_{l+1}} \omega(z) dz,\quad
    F_l^i = \int_{z_l}^{z_{l+1}} \bar \omega^i(z) dz, \quad
    i=2,\dots,L-2.
\end{equation}
Note that $F_l^i$ and $W_l^i$ form $(L-2)\times(L-2)$ matrices. In
Appendix, they are expressed in terms of multi-valued
hypergeometric functions \cite{Abel:2003yx,Opdam}. With
$SL(2,\C)$, we can set $z_1,z_{L-1},z_L$ to be $0,1,\infty$
respectively and the others to the cross-ratios of $x_i$.

Plugging (\ref{clansatz}), the solution is expressed as
\begin{equation}
  c_i F^i_l = f_{l+1} - f_l + v, \quad l = 1,\dots,L-2,
\end{equation}
where we defined $c_1 \equiv a, c_i\equiv b_i^*$, and
by inverting them we obtain
\begin{equation} \label{solglmonodromy}
  c_i = \sum_{l=1}^{L-2} (f_{l+1} - f_l + v)(F^{-1})^l_i ,
\end{equation}
where the inverse is taken with respect to the matrix basis with
indices $l,i$. Plugging into the classical action, we obtain the final
solution
\begin{equation} \label{claction}
  S_{\cl} (x_2,\dots,x_{L-2}) = \frac 1 {4\pi \alpha'} \left( |a|^2 I
  + \sum_{i,j} b^*_i b_j I'_{\bar \imath j} \right) ,
\end{equation}
where
\begin{equation}
  \begin{split}
    I(x_2,\dots,x_{L-2}) &= \int_{\C} d^2 z |\omega(z)|^2 ,\\
    I'_{\bar \imath j}(x_2,\dots,x_{L-2}) &= \int_{\C} d^2 z \bar
    \omega^i(\bar z)    (\bar \omega^j(\bar z))^*.
  \end{split}
\end{equation}
We can expand this action by products of
holomorphic and antiholomorphic functions, with careful choices of
contours. It is nothing but the
relation between open and closed string amplitudes
before integration over $x_i$s \cite{Kawai:1985xq},
\begin{equation} \begin{split} \label{openclosed}
  I(x_2,\dots,x_{L-2}) &= \sum_{i=2}^{L-1} (-1)^i \left[ 1-\exp(-2 \pi
  i \sum_{l=2}^i 
  \frac {k_l}{N}) \right] F_i \overline{F_1} \\
 &+ \sum_{j=2}^{L-2} \sum_{i=0}^{j-1}
  (-1)^{j+i+1} \left[ 1-\exp(- 2 \pi i
  \sum_{l=i+1}^j \frac{k_l}{N}) \right] F_i \overline{F_j},
\end{split} \end{equation}
where
\begin{equation} \begin{split}
  F_0    &\equiv \int_{-\infty}^0 dz \prod_{j=1}^{L-1} (z-z_j)^{-(1-k_j/N)}, \\
  F_i    &= F_i^1, \quad i=1,\dots,L-2, \\
  F_{L-1}&\equiv \int_1^\infty    dz \prod_{j=1}^{L-1} (z-z_j)^{-(1-k_j/N)}.
\end{split} \end{equation}

Plugging in (\ref{claction}), we obtain the classical action
$S_{\cl}(x_2,\dots,x_{L-2})$. 
It it a function of $L-3$ complex variables
$x_i$ which will be integrated out in the final amplitude.
Later, we will integrate this with variables $x_2, \dots x_{L-2}$ over
the entire complex plane. Among these,
using the saddle point approximation by adjusting $x_i$s or equivalently
ratios $F_{i+1}/F_i, i=2,\dots,L-2$, we find a minimum
\begin{equation} \label{mincond}
  { F_{i+1} \over F_{i}  } = {f_{i+2}-f_{i+1} \over f_{i+1}-f_{i}}.
\end{equation}
Note that in general the integrals $F_i$ are complex and we
coordinated the fixed points as complex vectors on a given 
2D torus and orbifold.

Inserting these into (\ref{solglmonodromy}), we have $c_i=b_i^*=0$
for all $i>1$, except $a=c_1 \ne 0$. The solution is
nothing but the generalized Schwarz--Christoffel transformation
\cite{Opdam}, whose original version maps the upper half plane into
inside an $L$-polygon
\begin{equation}
 \partial Z_{\cl} (z) = a \omega(z).
\end{equation}
Namely, the points $x_i$ are mapped to vertex $z_i$ and the turning
around angle is given by $k_i \pi /N$. In this case, we obtain the
instanton contribution is exponential of the polygon area
\begin{equation}
 {\cal Z}_{\cl,\rm min} \sim \exp \left[-\frac{1}{2 \pi \alpha'}
 \text{(``area of the polygon'')} \right].
\end{equation}
This is valid under the assumption (\ref{sumk}), i.e. forming a
polygon, but in general case we have more fundamental interpretation
shortly. 

Of course, there are other minima with the same value, where $a$
and all $b_i$ vanish except one, say $b_k \neq 0$. This corresponds
to $F_{i+1}^j/F_{i}^j = (f_{i+2}-f_{i+1})/(f_{i+1}-f_{i})$ and the
Schwarz--Christoffel transformation corresponds to $\bar \partial 
Z_{\cl} (\bar z) = b_k^* \bar \omega' (\bar z)^k$.

In forming the area from the Schwarz--Christoffel transformation,
the ordering is important. If we just exchange two fields, we
cannot satisfy the space group selection rule in general, and the
polygon becomes self-crossing, where the area rule is not
applicable. In the correlation function we take the radial ordering.
In the superpotential of effective field theory, 
we do not see the ordering, 
since  the integration over all $z_i$ completely
symmetrize the amplitude.

\subsection{Four-point correlation function}
The four-point correlation function provides a good example of
calculation of the classical part.
In this case, the functions $F_i$ are well-known
hypergeometric functions, which are solutions of second order linear
differential equation. It is known \cite{Whittaker-Watson}
that if any three of the solutions have the common domain of
existence, there be a linear relation among them. In our case, we
can express all of $F_i$ in terms of, say, $F_1$ and $F_2$.
They are shown in (\ref{f0}) and (\ref{f2prime}) of Appendix.

 Plugging these
to (\ref{openclosed}) the holomorphic part is obtained as
\begin{equation} \label{4ptclaction}
 I(x)= c_{11} |F_1|^2 + c_{12} F_1 \overline{F_2} + c_{12}^*
 \overline{F_1} F_2 + c_{22} |F_2|^2,
\end{equation}
where
\begin{align}
  c_{11} &= {\sin(\pi k_1/N) \sin(\pi(k_2+k_3)/N) \over \sin(\pi
  k_4/N)}, \nonumber \\
  c_{22} &= {\sin(\pi k_3/N) \sin(\pi(k_1+k_2)/N) \over \sin(\pi
  k_4/N)}, \\
  c_{12} &= -e^{\pi i  k_2/N} \left[ \sin (\pi k_2/N)
   + {\sin(\pi(k_1+k_2)/N)\sin(\pi(k_2+k_3)/N) \over \sin(\pi
     k_4/N)}\right].   \nonumber
\end{align}
The coefficient $c_{12}$ reduces to
\begin{equation}
c_{12} =  e^{\pi i k_2/N} \sin (\pi k_1/N) \sin(\pi k_3/N) ,
\end{equation}
only for the polygon case, using (\ref{sumk}).
The prefactor $e^{i \pi k_2/N}$ in $c_{12}$ is the
relative phase of (complex numbers) $v_{32}$ and $v_{21}$,
where $v_{ij}=f_i -f_j +v$ with $v \in \Lambda$.
{}From (\ref{solglmonodromy}) we have the
coefficients
\begin{equation} \begin{split}
  a &= { v_{32} \overline{F_1}' + v_{21} \overline{F_2}'
        \over F_1 \overline{F_2}' - F_2 \overline{F_1}' }, \\
  b &= { v_{32} F_1 - v_{21} F_2
        \over F_1 \overline{F_2}' - F_2 \overline{F_1}' }.
\end{split} \end{equation}
We can obtain the
antiholomorphic action $I'(\bar x)$ and integral $F'_i(\bar x)$ from
$I(x)$ and $F_i(x)$ respectively, by substituting $k_i/N \to
1-k_i/N$ and $x \to \bar x$. With these we obtain the classical
action (\ref{claction}). The action does not have manifest duality
symmetry, since we have fixed four points by $SL(2,{\bf C})$.

We define the following modulus
\begin{equation}
 \tau \equiv {F_2 \over F_1},
\end{equation}
which is in the $\Z_2$
case the modular parameter of two-torus, made by connecting two Riemann
sheets with two branch cuts \cite{Hamidi:1986vh,Dixon:1986qv}.
As expected from (\ref{mincond}), the minimum of $I(x)$ is obtained at
\begin{equation}
 \tau = {v_{32} \over v_{21}}.
\end{equation}
Thus we have $a=-v_{21},b=0$ and the minimum of classical action is
obtained as
\begin{equation}
  S_{\cl,\rm min} =  \frac{1}{2 \pi \alpha'} \Big[ c_{11} |v_{21}|^2
  + c_{12} v_{21} v_{32}^* + c_{12}^* v_{21}^* v_{32}
  + c_{22} |v_{32}|^2 \Big] .
\end{equation}
For the case of polygon, the classical action $S_{\cl,\rm min}$
reduces to
\begin{equation} \begin{split}
  S_{\cl,\rm min} =  & \frac{1}{2 \pi \alpha'} \left[
   {|v_{14}|^2 \over 2} {
   \sin (\pi k_1 /N) \sin (\pi k_4 /N) \over \sin(\pi (k_1+k_4)/N)
   } -
{|v_{32}|^2 \over 2} {
   \sin (\pi k_2 /N) \sin (\pi k_3 /N) \over \sin(\pi (k_2+k_3)/N)
   }    \right] \\
    = & \frac{1}{2 \pi \alpha'} \left[
   {|v_{43}|^2 \over 2} {
   \sin (\pi k_3 /N) \sin (\pi k_4 /N) \over \sin(\pi (k_3+k_4)/N)
   } -
{|v_{21}|^2 \over 2} {
   \sin (\pi k_1 /N) \sin (\pi k_2 /N) \over \sin(\pi (k_1+k_2)/N)
   }    \right].
\end{split} \end{equation}
This is the
{\em area of the quadrilateral} formed by vertices at the fixed
points $f_1,f_2,f_3$ and $f_4$.

In the case with $k_1+k_4 =N$ and/or $k_2+k_3=N$, this expression is not
well-defined.
Without loss of generality,
the case with $k_1+k_4 =N$ and $k_2+k_3=N$ leads to
$k_1 = k_3 = N-k, k_2 = k_4 = k$, by use of (\ref{sumk}), and
such a case has been  calculated in \cite{Dixon:1986qv}.
Here we have to come back to (\ref{4ptclaction}), and
the result agrees.\footnote{
To compare between our results and \cite{Dixon:1986qv},
we have to replace our modulus $\tau$ by $e^{\pi i
(k/N-1)}\tau$.}
The case with $ k_1+k_4 =N$ or $k_2+k_3=N$ leads to
$k_1=N-k,k_2=k,k_3=N-l,k_4=l$, and such a case
has been calculated in \cite{Burwick:1990tu}.

{}From four-point amplitude, we can obtain three point amplitude
by taking $x$ to, say,
$\infty$. In this case the fixed points $f_3$ and $f_4$ become
coalescent and the classical action $S_{\cl,\rm min}$ reduces to 
\begin{equation} \label{3ptclaction}
 S_{\cl,\rm min} = \frac{1}{2 \pi \alpha'} {|v_{32}|^2 \over 2} {
   \sin (\pi k_2 /N) \sin (\pi k_3 /N) \over \sin(\pi (k_2+k_3)/N)
   }.
\end{equation}
Note that this action depends on the choice of contour ``picture''.
Here, we chose one encircling two fixed points $f_3$ and $f_2$.
We do not need worry about whether $v$ is actually
compatible to factorization \cite{Dixon:1986qv}. The Pochhammer loops
in which $v_{32}$ and $v_{14}$ belong are independent.

In the special case with $k_1=k_2=k_3=k_4=N/2$, we have $c_{11}=c_{22}=0,
c_{12} =-i$ and $F_1'=F_1, F_2'=-F_2$ yielding
\begin{equation} \label{clrectangle}
 S_{\cl} = \frac{1}{2\pi\alpha
'} 2 v_{21} \frac{v_{32}}{i} = \frac2{2\pi \alpha'} |v_{21}v_{32}|,
\end{equation}
which is again interpreted as {\em twice} the area of the rectangle,
in unit of $2 \pi \alpha'$ if $v_{21}$ is orthogonal to $v_{32}$.
This is the case of order 2 subsector ($N/2$-th twisted sector) in
even order orbifold. Note that this action is not the minimum action,
since in this case the classical action is not the function of
$\tau$. In this case the area rule interpretation is somehow
ambiguous. We will study more detail in the following subsection.

We have considered $L$-point couplings only for $L=2,3,4$ as
examples. However, we will study that higher order $L$-couplings
reduce to a combination of lower order $L'$-couplings with $L' <
L$ by the discussion of field coalesce in subsection 3.4. In
addition, since the number of fixed points on $T^2/\Z_N$ orbifolds
is limited, we can expect that most of higher order $L$-couplings can
be written as combinations of $L$-point couplings only with
$L=2,3,4$. We will study this expectation in separated papers
\cite{CKNRV,CK}, by examining concrete orbifold models.

\subsection{Non-polygon case: the meaning of area}

We assumed the polygon condition (\ref{sumk}) is satisfied.
However, in general,  the following relation
\begin{equation}
 \sum_{i=1}^L \frac{k_i}{N} \le L-2,
\end{equation}
is possible.
In the inequality case, the holomorphic part of the classical
solution decays faster than $z^{-2}$, whereas the antiholomorphic
part decays not faster than $z^{-2}$. The only sensible way of
treating is to make antiholomorphic part vanishing.

\begin{figure}[t]
\begin{center}
\includegraphics[height=5cm]{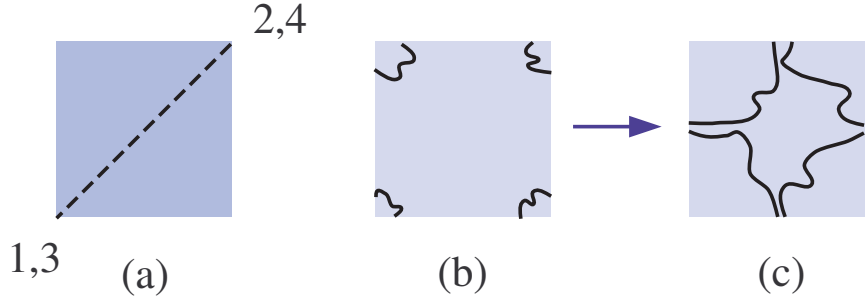}
\caption{(a) On $\Z_4$ orbifold, four order 4 couplings: two at a
vertex, and the other two at the opposite vertex. (b) Massless
strings are localized (c) Quantum effect grows twisted strings to
form a untwisted string, which is a intermediate state described by
instanton. The shaded area, where $v_{21}$ is the diagonal, is the
minimal area swept.} \label{fig:swptarea}
\end{center}
\end{figure}
\begin{figure}[t]
\begin{center}
\includegraphics[height=3.5cm]{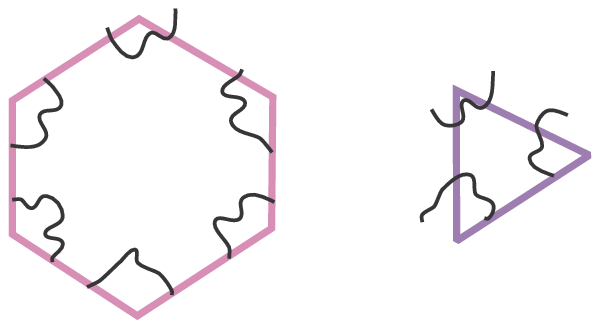}
\includegraphics[height=3.5cm]{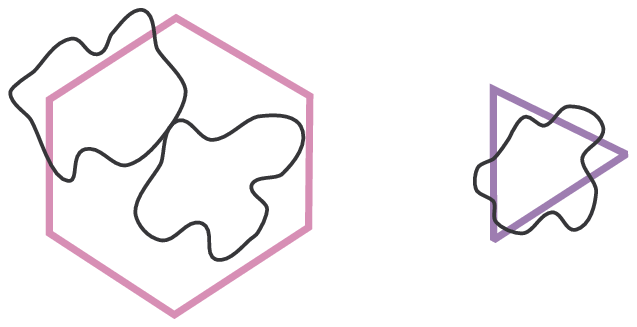}
\caption{Naive area rule is the special case of the swept-are rule
  when the twisted strings form a polygon.}
\end{center}
\end{figure}

For example, in the $\Z_4$ orbifold, the coupling of four first
twisted sector fields,
$$ \sigma_{(\theta,0)}\sigma_{(\theta,e_1)}\sigma_{(\theta,0)}
\sigma_{(\theta,e_1)}, $$
shown in Fig. \ref{fig:swptarea}(a), satisfies the
space group selection rule.
Note that
$\sigma_{(\theta,0)}\sigma_{(\theta,0)}\sigma_{(\theta,e_1)}
\sigma_{(\theta,e_1)}$
does not satisfy the space group selection rule, and
the ordering is important.
All of twist fields are of order four, which cannot satisfy the relation
(\ref{sumk}).
The classical solution is obtained as
\begin{equation} \begin{split}
 \partial Z_{\cl} &= a(z-z_1)^{-1/4} (z-z_2)^{-1/4} (z-z_3)^{-1/4}
 (z- z_4)^{-1/4}, \\
 \bar \partial Z_{\cl} &= 0.
\end{split} \end{equation}
{}From global monodromy condition, we have
$$ a = \frac{v_{21}}{F_1},\quad \frac{F_2}{F_1} = \frac{1}{\sqrt 2}(-1 +
 i) ,$$
and the classical action is given with
$c_{11}=c_{22}=1,c_{12}=-\frac32(1+i)$, yielding
$$ I = 2. $$
Thus we have the classical contribution
$$ S_{\cl} = \frac{1}{2 \pi \alpha'} 2 |v_{21}|^2,  $$
which is interpreted as the area of the square whose diagonal is
$v_{21}$.

Similarly, for the coupling
$$ \sigma_{(\theta,0)}\sigma_{(\theta^2,e_1)}\sigma_{(\theta,e_1)}, $$
we obtain the classical action,
$$ S_{\cl} = \frac{1}{2 \pi \alpha'} \frac12 |v_{21}|^2. $$

In these cases, we have a different area rule: The ``area'' is not
that surrounded by fixed points, but one as follows. The classical
solution describes a local minimum of the action, which is the
instanton of worldsheet nature, suppressed by $\alpha'$. The
selection rule tells us that these twisted strings can potentially
make an untwisted string, which is not possible due to energetics
for massless strings, since they are completely localized at certain
fixed points, as in Fig. \ref{fig:swptarea}(b). However they can
oscillate to grow to be large size, and above a certain threshold,
they can form an untwisted string as in Fig \ref{fig:swptarea}(c).
Noting that the instanton describes tunneling between vacua which is
energetically forbidden, we can understand that forming untwisted
string corresponds to such tunneling.

Still we can have the hint from the modified area. It is the {\em
sweeping area for localized twisted strings to grow to become a
untwisted string.} For the polygon case, i.e. that satisfying the
condition (\ref{sumk}), this interpretation is still valid, since
still the area swept by twisted strings at each vertex makes the
polygon area.

On the other hand, in the case not satisfying the condition
(\ref{sumk}),
we lost the interpretation of the mapping $\partial Z$
being a generalized Schwarz--Christoffel transformation, since in the
target space the fixed points fail to make a polygon.

The subleading correction is generated by identical fixed points on
the orbifold, but more separated in the covering space, i.e.
$f+(1-\theta^k)\Lambda$. Since
they are identical points, they satisfy the selection rule.
With this interpretation, we can understand the classical solution
which makes more than one untwisted strings possible.
For example, for $\sigma_{(\theta,0)}\sigma_{(\theta^2,e_1)}
\sigma_{(\theta,e_1)}$, the coupling
$\sigma_{(\theta,0)}\sigma_{(\theta^2,3e_1)}
\sigma_{(\theta,3e_1)}$, where each fixed point belongs
to the same conjugacy class as the previous one, satisfies
the space group selection rule, but corresponds to a large
$v_{21}$ and a large instanton action.

\subsection{Fields coalescent at the same fixed point}

In most cases, some of the fields sit at the same fixed point $f$. 
The corresponding correlation function might be obtained 
by taking the limit $ z_j \to z_i$ in the correlation 
function, 
\begin{equation} \label{coallimit}
 \langle \dots \sigma_{f,k_1} (z_i) \dots \sigma_{f,k_2} (z_j) \dots
 \rangle  .
\end{equation}
Note that this limit is not always well-defined. It is because 
twist fields do not commute.
By conformal symmetry, the OPE has the generic form,  
\begin{equation} \label{OPEsigmas}
 \sigma_{f,k_i} (z_i) \sigma_{f,k_j} (z_j) \sim c_{ij} (z_i-z_j)^{\kappa_{ij}}
 \sigma_{f,k_i+k_j} (z_j).
\end{equation}
Equating the conformal weights of the both sides, we have
\begin{equation}
  \kappa_{ij}=h_{\sigma_{k_i+k_j}}-h_{\sigma_{k_i}}-h_{\sigma_{k_j}} =
 \left\{ \begin{matrix} -{k_i \over N} {k_j \over N} & (k_i+k_j \le N) \\
-(1- {k_i  \over  N})(1-{k_j \over N}) & (k_i+k_j \ge N) \\
  \end{matrix} \right. .
\end{equation} 
Because of the nontrivial branch cut, this relation is asymmetric
under the exchange of two twist fields. This property is also
reflected in the space group elements, which do not commute, either.
We can define an invariant block of twist fields, which correspond to 
the identity of the space group $(1,0)$.
Thus, these invariant blocks commute and satisfy 
\begin{equation}
  \sum_j \kappa_{ij} = 0 \mod 1.
\label{condition}
\end{equation}

In the case where two points $z_i$ and $z_{i+1}$ are in the
successive order, and we can merge two twists without ruining the
radial ordering. They are {\em neighboring} points as polygon vertices.
Then, from (\ref{OPEsigmas}) the two twists reduce to a single twist
with the summed order. Also this implies the classical solution becomes
\begin{equation}
 \cdots  (z - z_i)^{k_i/N-1} (z - z_{i+1})^{k_{i+1}/N-1} \cdots \to
 \cdots  (z-z_i)^{(k_i+k_{i+1})/N-2} \cdots
\end{equation}
where, again, the even integral power is not relevant to branch
structure, so that we can make $k_i/N + k_{i+1}/N -1$ lie in $[-1,1)$
for instance.
This means, not all of $L$ vertices form the $L$-polygon, but effectively
one with the lesser vertices. (Recall that the positions of vertices
are given as singularities in the classical solution.)
In fact this is the familiar case when
we obtain a three point function from the four point function by
setting two of the points coalesce. In the latter limit, the polygon is
triangle.
We can see this in terms of space group elements. Neglecting gauge
group, which is not involved in the classical amplitude, we cannot
distinguish the product
\begin{equation}
 (\omega,v)^2 \text{ and } (\omega^2,v+ \omega v),
\end{equation}
when two identical twisted fields sit at the same point, and are
put on the neighboring points in the correlation function.

There is the case where $k_i + k_{i+1}$ is integer. In
this case there is no singularity in the classical solution
and such vertex does not contribute to the area rule.
In the extreme case where $k$ order-$k$ twisted fields sit at
the same points, the coefficients of higher order couplings are
not suppressed. Thus all we need to consider is the other
nontrivial couplings. Fortunately, not all of them survive: From
$R$-symmetry invariance, the correlation function is further
constrained. 

\begin{figure}[t]
\begin{center}
\includegraphics[height=6cm]{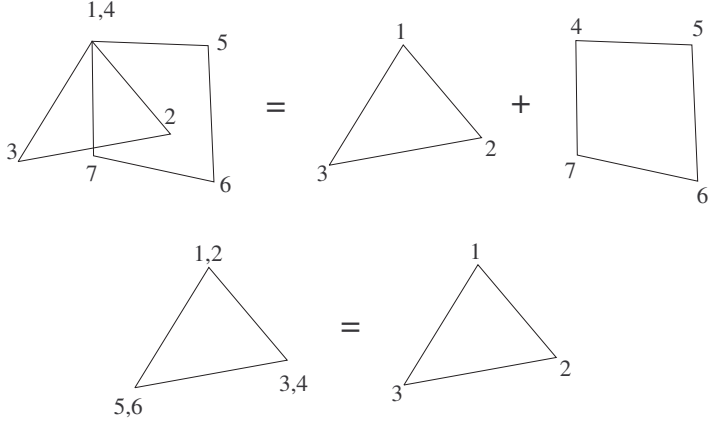}
\caption{The rule for worldsheet instanton sum.} \label{fig:wsinst}
\end{center}
\end{figure}

However, from the radial ordering, there is a case in which
the exchange of two branch cut fails to give well-defined radial
ordering.
This is not possible for two fields which are not the neighboring
fields in the correlation function, since the space group and the
mapping of classical solution is not commutative.

For example, in couplings among twist fields corresponding to
two $(\theta,0)$'s, two $(\theta,e_1)$'s and two
$(\theta,e_1+e_2)$'s on the $\Z_3$ orbifold,
there are two following combinations
possible, satisfying the space group selection rule
\begin{align}
 \sigma_{(\theta,0)} & \sigma_{(\theta,0)} \sigma_{(\theta,e_1)}
    \sigma_{(\theta,e_1)} \sigma_{(\theta,e_1+e_2)}
    \sigma_{(\theta,e_1+e_2)}, \\
 \sigma_{(\theta,0)} & \sigma_{(\theta,e_1)} \sigma_{(\theta,e_1+e_2)}
    \sigma_{(\theta,0)} \sigma_{(\theta,e_1)}
    \sigma_{(\theta,e_1+e_2)}.
\label{notreducible}
\end{align}
In the first case, the solution behaves like
\begin{equation} 
 \partial Z  =
 a(z-z_1)^{-2/3}(z-z_2)^{-2/3}(z-z_3)^{-2/3}
           (z-z_4)^{-2/3}(z-z_5)^{-2/3}(z-z_6)^{-2/3}.
\end{equation}
The two points, $z_i$ and $z_{i+1}$ for $i=1,3,5$, correspond 
to the same fixed points on the target space, i.e. 
$Z(z_1)=Z(z_2)=f_1$, $Z(z_3)=Z(z_4)=f_2$ and $Z(z_5)=Z(z_6)=f_3$.
Thus, we take the limit $z_{i+1} \rightarrow z_i$ for $i=1,3,5$.
In such a limit, the solution behaves like 
\begin{equation} 
\partial Z    =  a(z-z_1)^{-4/3}(z-z_3)^{-4/3}(z-z_5)^{-4/3}  .
\end{equation}
Thus, in the space group point of view, it is indistinguishable from the
coupling among the second twisted couplings
\begin{equation}
 \sigma_{(\theta^2,0)} \sigma_{(\theta^2,e_1+e_2)}
 \sigma_{(\theta^2,e_2)} ,
\end{equation}
which gives only the area of triangle.
On the other hand, the latter coupling (\ref{notreducible}) gives
{\em twice the area} of the triangle. The classical solution maps
from the different points to the same points. For example the
holomorphic part behaves
\begin{equation} 
 \partial Z =
 a(z-z_1)^{-2/3}(z-z_2)^{-2/3}(z-z_3)^{-2/3}(z-z_4)^{-2/3}(z-z_5)^{-2/3}(z-z_6)^{-2/3}, 
\end{equation}
and the two points, $z_i$ and $z_{i+3}$ for $i=1,2,3$, correspond 
to the same fixed points on the target space, i.e. 
$ Z(z_1)=Z(z_4)=f_1$, $Z(z_2)=Z(z_5)=f_2$ and $Z(z_3)=Z(z_6)=f_3$.
Otherwise the ordering is ruined. It is also understood that two
different and {\em independent} sets of twisted fields can sweep the
triangle, which means the instanton corrections from the complete
polygons are additive in the action. This case is depicted in Fig.
\ref{fig:wsinst}.

We summarize how to calculate the classical part of $L$-point
couplings among twist fields $\sigma_{k_i}$ ($i=1,\cdots, L$).
First, we classify all of possible ordering of these twist fields,
which satisfy the space group selection rule. For each ordering of
twist fields allowed by the space group selection rule, we
consider the following procedure. We combine two or more twist
fields sitting at the same fixed point to a single twist field
like (\ref{OPEsigmas}) and Fig.~ 3 if possible, that is, they
satisfy (\ref{condition}). When their total twist is just $(2 \pi
n)$ twist with integer $n$, correlation function reduces to much
simpler form. The resultant correlation function can be written as
product of invariant blocks, which satisfy the space group selection rule
like Fig.~3. Each block includes smaller number $L'$ of twist
fields.\footnote{ In most of cases, $L'$ may be equal to
$L'=2,3,4$. We will study this point in concrete models
\cite{CK}.} Then, for each block, we calculate classical
contributions, $e^{-S_{\rm cl}}$, i.e. instanton actions
corresponding to the minimum action and larger ones, and take
their summation, i.e. $\sum_{i,\{v_{i+1,i}\}} \exp(-S_{\cl})$.
Next, we take a production of classical contributions
corresponding to each  block, i.e.
\begin{equation} \label{sumrule}
 \prod_{\text{invariant blocks}} 
\left( \sum_{i,\{v_{i+1,i}\}} \exp(-S_{\cl}) \right).
\end{equation}
Finally, we sum over all of possible ordering to obtain 
the total coupling, i.e,
\begin{equation} \label{sumrule}
\sum_{\text{possible ordering}} \left[ \prod_{\text{invariant blocks}} 
\left( \sum_{i,\{v_{i+1,i}\}} \exp(-S_{\cl}) \right) \right].
\end{equation}

Because of the ordering, the correlation function does not possess
the worldsheet duality like in the Virasoro--Shapiro amplitude
\cite{ViSa}. 
In the vanishing momentum limit we do not distinguish the channel,
thus the effective coupling like Yukawa coupling does not
distinguish the order. However even in this case, the above two
cases are distinguishable, since this contribution is the
worldsheet effect, suppressed by $\alpha'$.

\subsection{Linearly combined states}
In higher order twisted sector of a non-prime order orbifold, there are
states formed by linear combinations, as in (\ref{lincombi}) , due to
the orbifold projection.  
The linear combination of states can be more precisely defined by that of
vertex operators. It follows that the corresponding classical solution
consists of linear combination of individual solutions before 
combination, and also the relative weights are inherited.
For example the classical solution involving (\ref{lincombi})
contains the factor
\begin{equation}
  \frac{1}{\sqrt k}\bigg( 
 (z-  z_1)^{-1+k/N} + \gamma ( z- z_2)^{-1+k/N}
+ \dots + \gamma^{k-1} (z- z_k)^{-1+k/N} \bigg),
\end{equation}
where $z_i$ is mapped to the fixed points $(\theta,\theta^{i-1}v)$.

For such couplings, we observe two points
\begin{enumerate}
 \item The selection rule for a linearly combined state is derived from
 that for each term. 
 \item A part of the classical solution, which does not satisfy the
 selection rule vanish.
\end{enumerate}
The proof for the first is given in Ref. \cite{CKNRV}, and
we easily see the latter is the case. We can show that such term
not satisfying the classical solution always contain the same element
more than once. Setting these to the other(s) corresponding to
shrinking the area to zero, if we do not want to change the angles
corresponding other vertices. 

For instance, in the second twisted sector of $\Z_4$ orbifold, we have a
coupling including linearly combined states
\begin{equation}
 \sigma_{(\theta^2,0)} \cdot \frac{1}{\sqrt2} \left(\sigma_{(\theta^2,e_1)} \pm
 \sigma_{(\theta^2,e_2)} \right) \cdot \sigma_{(\theta^2,e_1+e_2)} \cdot
 \frac{1}{\sqrt2} \left(\sigma_{(\theta^2,e_1)} \pm
 \sigma_{(\theta^2,e_2)} \right).
\end{equation}
The total coupling is given by summation over 
each of the four correlation functions. However, the nonvanishing
contributions come from the only ones satisfying the selection
rule. One that does not satisfy the rule, e.g. $\sigma_{(\theta^2,0)}
\sigma_{(\theta^2,e_1)}\sigma_{(\theta^2,e_1+e_2)}\sigma_{(\theta^2,e_1)}$, 
does not contribute. 

To sum up, for the linearly combined state, we can treat a linearly
combined state as a complete physical field, the classical amplitude
has only contribution from the parts satisfying the space group
selection rule.

\section{The quantum amplitude}

Now we determine the quantum part of the amplitude (\ref{corrclandqu}),
\begin{equation} \label{qntpt}
 {\cal Z}_{\qu} = \langle \sigma_{k_1} \dots \sigma_{k_L} \rangle_{\qu}.
\end{equation}
We use the stress
tensor method \cite{Dixon:1986qv,Burwick:1990tu,Abel:2003yx}, in
which (\ref{qntpt}) can be indirectly calculated, relying on only the
holomorphicity. All the information can be read from the Green's function
\begin{equation} \label{grnfn}
  g(z,w;z_i) \equiv { \langle -\frac12 \partial_z Z \partial_w
  \overline Z \prod \sigma_{k_i} \rangle \over {\cal Z}_{\qu}}.
\end{equation}
Its form we know from the holomorphic part of the OPE (\ref{OPEdZsigma1}),
\begin{equation} \label{holgreen}
  g(z,w) = \omega(z) \omega'(w) \left [ \sum_{i<j} a_{ij} {(z-x_i)(z-x_j)
  \prod_{k\ne i,j} (w-x_k) \over (z-w)^2} + A(w) \right ],
\end{equation}
where $\omega'(w) = \prod (w-z_i)^{-k_i/N}$.
The condition $i<j$ avoids overcounting, although otherwise the
formula would be more symmetric.
The prefactor $\omega(z) \omega'(w)$, from (\ref{omegas}), gives the
desired pole structure. The normalization is determined by 
requiring the $z \to w$ limit of $g(z,w)$ to be 1, and the condition
\begin{equation} \label{doublepolecoeff}
 \sum_{i<j} a_{ij} =1 ,
\end{equation}
leads the desired conformal weight $h_k$, as the coefficient of
each double pole in $z-w$.
To have no residue in $z-z_i$, we require the condition
for $a_{ij}$,
\begin{equation} \label{aijcond}
  \sum_{j=1}^{L} a_{ij} = 1- {k_i \over N},
\end{equation}
where we use the fake number $a_{ij}=a_{ji}$ for the case $i>j$. Summing
over $i$, this condition implies (\ref{doublepolecoeff}).
We have less condition for many $a_{ij}$, reflecting the freedom of
our choice of $A$. Of course, the physics does not depend on the choice
$a_{ij}$. As $z \to w$, we have the OPE
\begin{equation} \label{TsigmaOPE} \begin{split}
  {\langle T(z) \prod \sigma_{k_i} \rangle \over
{\cal Z}_{\qu}}  \sim  -& \frac12 \sum_{i<j}
\frac{k_i}{N} \frac  {k_j}{N} \frac{1}{(z-z_i)(z-z_j)} \\
& + \frac12 \sum_{i<j} a_{ij} \left(
  \frac{1}{z-z_i} + \frac{1}{z-z_j} \right)^2 + { A \over \prod_i
  (z-z_i)} .
\end{split} \end{equation}
We extract the residue $(z-z_k)$ for fixed $k$, and take $z \to
z_k$ limit.

In the limit $w \to z$, the Green's function becomes
energy-momentum (stress) tensor
\begin{equation} \label{greenpole}
  T(z) \sim \lim_{w \to z} \left[- \frac12 \partial_z Z \partial_w
  \overline Z -  \frac{1}{(z-w)^2} \right],
\end{equation}
where the last term arises from the normal ordering. Sandwiching
the OPE in the correlation function
\begin{equation} \label{residuediff}
  T(z) \sigma_{k_k} (z_k) \sim {h_k \sigma_{k_k}(z_k) \over (z-z_k)^2} +
  {\partial_{z_k} \sigma_{k_k} (z_k) \over z - z_k},
\end{equation}
with the fixed $\sigma_{k_k}(z_k)$ and its conformal weight $h_k$
given in Eq. (\ref{confwght}),
we can completely calculate the holomorphic part ${\cal Z}^{\h}_{\qu}$ up
to normalization
\begin{equation}
  \partial_{z_k} \ln {\cal Z}^{\h}_{\qu} 
  =    { \partial_{z_k}  {\cal Z}^{\h}_{\qu} \over {\cal Z}^{\h}_{\qu}} =
\lim_{z \to z_k} (z-z_k)
  \left[  {\langle T(z) \prod \sigma_{k_i} \rangle_{\qu} \over {\cal
  Z}^{\h}_{\qu}}  - {h_k \over  (z-z_k)^2} \right].
\end{equation}
We obtain
\begin{equation}
  \partial_{z_k} \ln {\cal Z}^{\h}_{\qu} = \partial_{z_k} \sum_{i<j} \left [a_{ij} -
\left(1-\frac{k_i}{N} \right) \left( 1-\frac{k_j}{N}\right)
 \right]{ 1 \over z_k - z_j} + {A \over \prod_{i \ne k} (z_k-z_i)}.
\end{equation}

At this stage, using $SL(2,\C)$ symmetry, we set $z_1 \to 0, z_2
\to x, z_{L-1} \to 1, z_L \to \infty$. Then $z_L$ dependent terms
vanish since $z_L \to \infty$, but there are corrections to this
from $A$, which are also dependent on $z_L$. We now have $L$
unknown for $L-1$ constraint,\ and remaining freedom corresponds
to redefinition of $A$.

Similarly we define the antiholomorphic Green's function
\begin{equation}
  h(\bar z,w) = \sum_{i,j=2}^{L-2} B_{ij} \omega^{\prime i} (\bar z)
  \omega'_j (w).
\end{equation}
To obtain $A$ and $B_{ij}$, we apply the global monodromy
condition for the quantum part
\begin{equation} \label{glmoquant}
  \oint_{C_l} dz g(z,w) + \oint_{C_l} d \bar z h(\bar z,w) = 0 .
\end{equation}
It is convenient to call the first term in the 
RHS (\ref{holgreen}) as $g_s(z,w)$. If all the $z_i$ are
different, the matrix $W_i^l$ has the inverse. Thus we can
multiply it to eliminate $A$ and $B_{ij}$, so that
we obtain 
\begin{equation} \begin{split}
 \omega'(w) A &= - \sum_{l}^{L-2} (W^{-1})_1^l \oint_{C_l}
 dy g_s(y,w), \\
 \sum_{j=1}^{L-2} \omega^{\prime j} (w) B_{ij} &= - \sum_l^{L-2}
 (W^{-1})^l_i \oint_{C_l} dy g_s(y,w).
\end{split} \end{equation}
Since we set $z_L \to \infty$, the only relevant terms are ones
containing the factor $a_{iL} (z-z_L)$. We divide them by
$\omega'(w)$ and take the limit  $w \to \infty$.  Then, in the
sense of (\ref{residuediff}), we read off the residue of $(z-z_k)$
in the integrand
\begin{equation}
\lim_{z \to z_k} (z-z_k) \left[ \lim_{w \to \infty} {g_s(z,w) \over \omega'(w)}
    \right]
 \sim \partial_{z_k} \omega(z)
 - \sum_{i < j} a_{ij} \frac{1}{z_k-z_j}
 + \sum_{j \ne k} \left(1-\frac{k_k}{N} \right) {1 \over z_k - z_j} ,
\end{equation}
where we have replaced $a_{iL}$ with $a_{ij}$ by the relation
(\ref{aijcond}).
Therefore we have no dependence on the specific choice of $g_s(z,w)$,
or $a_{ij}$, as it should be.

Integration around the contour $C_l$ gives
\begin{equation} \label{zkzqu} \begin{split}
  \partial_{z_k} \ln {\cal Z}^{\h}_{\qu} = & -\sum_l (W^{-1})^l_1
  \partial_{z_k} W^1_l \\
 & + \partial_{z_k} \ln \left[ \prod_{j \ne k}
  (z_k - z_j)^{(1-\frac{k_k}{N})} \prod_{i \ne j} (z_i -
  z_j)^{-(1-\frac{k_i}{N})(1-\frac{k_j}{N})} \right] .
\end{split} \end{equation}
The first term in the RHS is contained in the derivative of $\det W \equiv \det
W_l^k$. By the chain rule, we obtain
\begin{equation} \label{wonderfulid}
 \partial_{z_k} \ln (\det W) =  \sum_{l=1}^{L} (W^{-1})^l_i
  \partial_{z_k} W^i_l + \sum_{j=1,j \ne k}^{L-3} \sum_{l=1}^{L} (W^{-1})_j^l
   \partial_{z_k} W^j_l.
\end{equation}
For the second term in the RHS of (\ref{zkzqu}), in the same way, we find the
singular structure as $z \to z_k$,
\begin{equation}
  \sum_{j \ne k} \sum_{l=1}^{L-2} (W^{-1})^l_j \partial_{z_k} W^j_l =
 \partial_{z_k} \ln \prod_{j \ne k} (z_j-z_i)^{\frac{k_i}{N}}.
\end{equation}
Thus we obtain
 \begin{equation}
  \partial_{z_k} \ln {\cal Z}^{\h}_{\qu} = \partial_{z_k} (\det W)^{-1} \ln
 \left[ \prod_{i,j=1,i<j}^{L-3} (z_i - z_j) \prod_{i<j=1}^{L} (z_{i} -
 z_{j})^{-(1-\frac{k_{i}}{N})(1-\frac{k_{j}}{N})} \right] ,
\end{equation}
up to antiholomorphic action. In fact, there are additional factors,
 since we have used only the part of the terms $\sum_l
 (W^{-1})^l_1 \partial_{z_k} W^1_l$ in (\ref{wonderfulid}). The
 remaining terms completely vanish if we calculate the antiholomorphic
 part in the same way.
The similar analysis is carried out for $B_{ij}$,
by dividing by $\omega^{\prime i}(w)$.
Combining it with the holomorphic part ${\cal Z}_{\qu} = {\cal
 Z}^{\h}_{\qu} \cdot {\cal Z}^{\rm antih}_{\qu}$, we arrive
\begin{equation} \label{ampuptonorm}
  {\cal Z}_{\qu} = (\det W)^{-1}
 \prod_{i,j=1,i<j}^{L-2}  (\bar z_i - \bar z_j)
 \prod_{i,j=1,i<j}^L (z_i -  z_j)^{-(1-\frac{k_i}{N})(1-
\frac{k_j}{N})} (\bar z_i - \bar z_j)^{-\frac{k_i}{N}\frac{k_j}{N}},
\end{equation}
up to the normalization, to be determined in the following section. 
The various branch cuts reflect the noncommutativity of the vertex
operators under permutation, due to orbifold phase. As well as the
classical part, the quantum part
has no world-sheet duality, 
either. Nonetheless the complete amplitude will 
be single-valued, completed with other parts of vertex operators.

\section{Factorization and normalization}

To normalize the amplitude, we choose the reference of
normalization
\begin{equation} \label{normalization}
 \langle \sigma_k(z,\bar z) \sigma_{N-k}(0,0) \rangle = 1 \cdot
 |z|^{-2\frac{k}{N}(1-\frac{k}{N})}, \quad z \to 0.
\end{equation}
Equivalently, two twist fields of the opposite twists coalescent
at the same point become the identity operator. In general we
cannot make use of the normalization (\ref{normalization}), since
for each twist field we need one with the opposite twist. Thus we
use a doubling trick.

\begin{figure}[t]
\begin{center}
\includegraphics[height=3cm]{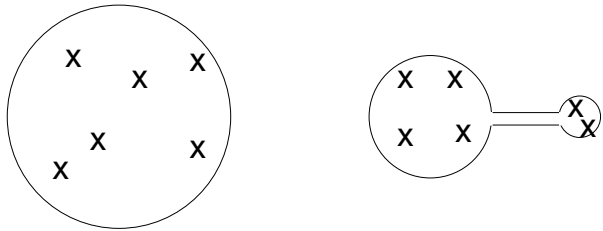}
\caption{Factorization of $2(L-1)$-point amplitude to
$[2(L-1)-k]$-point and $k$-point amplitudes. Unitarity tells us the
intermediate states have mass poles, whose residues are the product of
the factor amplitudes. The correct
normalization does not depend on a specific factorization and choice
of intermediate state.}
\end{center}
\label{fig:factorization}
\end{figure}

First we calculate $2(L-1)$ point function with
twists
\begin{equation}
S(1;2;\dots;2L-2) =  \langle \sigma_{k_1}(z_1) \sigma_{N-k_1}(w_1) \dots
 \sigma_{k_{L-1}}(z_{L-1}) \sigma_{N-k_{L-1}}(w_{L-1})
\rangle.
\end{equation}
By setting $w_{L-1} \to z_{L-1}$ we obtain the product of a
$(2L-3)$-point function and a $3$-point function. From the unitarity,
the intermediate state gives an on-shell pole, with the lowest
state being massless gauge bosons,
\begin{equation}
  S(1;2;\dots;2L-2) = i \int {d^4 k \over (2\pi)^4} {S(1;2;\dots;2L-4;j)
  S(j;2L-3;2L-2) \over -k^2 + i \epsilon }.
\end{equation}
This factorization is schematically drawn in Fig.
\ref{fig:factorization}.

Concentrating only twist fields, the intermediate gauge boson does
not contain a twist field. The former contains only $2L-4$ twist
operators and the latter has the form as in (\ref{normalization}).
Therefore we have
\begin{equation}
 \langle \sigma_{k_1} \sigma_{N-k_1} \dots \sigma_{k_{L-1}} \sigma_{N-k_{L-1}} \rangle
   =
 \langle \sigma_{k_1} \sigma_{N-k_1} \dots \sigma_{k_{L-2}} \sigma_{N-k_{L-2}} \rangle
  \langle
  \sigma_{k_{L-1}} \sigma_{N-k_{L-1}}
  \rangle.
\end{equation}
We have already known the general solution. For this we have the same
action, including the classical and the quantum parts, except the
factor $\det W$. In this, each term $\prod_l W_l^{i_l}$ contains the
common factor
\begin{equation} \label{commfactor}
 \prod_{l=1}^{L-2} \bigg(4 e^{i(k_{i+1}-k_i)\pi/N} \sin
 \frac{k_{i+1}\pi}{N} \sin \frac{k_{i}\pi}{N} \bigg).
\end{equation}
Thus there is a discrepancy between the two by the factor
\begin{equation}
 {\det W_{(l)} \over \det W_{(l-1)}} =4 e^{i(k_l-k_{l-1})\pi/N} \sin
 \frac{k_{l}\pi}{N} \sin \frac{k_{l-1}\pi}{N} ,
\end{equation}
where the subscript in the determinant indicates that these are for $l$ and
$l-1$ point correlation functions, respectively. This is so, since the
normalization of $2(L-1)$ point function have no special dependence
of a specific $k_i$. In effect, we replace $\det W$ in
(\ref{ampuptonorm}) with $\det F$, which is defined in
(\ref{Fmatrix}).
This is a desirable result, since any choice
of factorization should give the same result, not depending on the
special set of contour choice, which is reflected in $\sin
\alpha_i$.

We cannot go  from the four-point function into the product of two
two-point functions, since the point $z$ cannot be arbitrary close
to all of $0,1,\infty$, already fixed by $SL(2,\C)$, at the same
time. For this, we do the Poisson resummation \cite{Dixon:1986qv}.
Since it involves a lattice transformation into its dual lattice,
as a result we have the overall factor of the lattice volume.

Finally we set $z_L, z_{L+1}, \dots z_{2L-2}$ to infinity to obtain
the OPE
\begin{equation}
  \langle \sigma_{k_1} \dots \sigma_{k_L} \rangle
   \sim \sum_{(\theta^k,v) \in {\sf P},\rm sel} \langle
   \sigma_{k_1} \dots \sigma_{k_{L-1}}
  \sigma_{-\sum_1^{L-1}k_i} \rangle
 \langle \sigma_{\sum_1^{L-1}k_i}
  \sigma_{k_{i+1}} \dots \sigma_{k_L} \rangle ,
\end{equation}
where the selection rules should be satisfied. The RHS is nothing
but the product of two identical $(L-1)$th order couplings.
Comparing the coefficients, we obtain the coupling of order $L$
\begin{equation} \label{Lptqtsize}
  Y_{L\text{-point}}(k_1,k_2,\dots,k_L) = \lim_{\text{all } w_i \to
  \infty} \big[ \det
  F_{(2L-2)}(k_1,k_2,\dots,k_{L-1},N-k_1,N-k_2,\dots,N-k_{L-1})
  \big]^{-\frac12},
\end{equation}
where
\begin{equation}
 \frac{k_L}{N} =L-1-\sum_{j=1}^{L-1} \frac{k_i}N.
\end{equation}

For the rest of vertex operator components, we have
\begin{align}
 \langle \tilde c(z_1) c(\bar z_1) \tilde c(z_2) c (\bar z_2) \tilde
 c(z_3) c(\bar z_3) \rangle &=
 |z_1-z_2|^2 |z_1-z_3|^2 |z_2-z_3|^2,
\nonumber \\
  \langle e^{-\phi} (\bar z_1) e^{-\frac12 \phi} (\bar z_2) e^{-\frac12 \phi}
  (\bar z_3) \rangle &= (\bar z_1-\bar z_2)^{-1/2}(\bar z_1-\bar
  z_3)^{-1/2}(\bar z_2-\bar z_3)^{-1/4}, \nonumber \\
  \langle \psi^\mu (\bar z_2) \psi^\nu (\bar z_3) \rangle &= \eta^{\mu
  \nu}(\bar z_2-\bar z_3)^{-1}, \nonumber \\
 \left \langle \prod e^{i k_i \cdot X}(z_i, \bar z_i) \right \rangle & =
  \prod_{i,j=1,i<j}^{L} |z_i-z_j|^{\alpha' k_i \cdot k_j}, \\
 \left \langle \prod e^{i P_i \cdot X}(z_i) \right \rangle & =
  \prod_{i,j=1,i<j}^{L} (z_i-z_j)^{\alpha' P_i \cdot P_j/2},
\nonumber  \\
 \left \langle \prod e^{i r_i \cdot H}(\bar z_i) \right \rangle & =
  \prod_{i,j=1,i<j}^{L} (\bar z_i-\bar z_j)^{\alpha' r_i \cdot r_j
    /2}. \nonumber
\end{align}
Multiplying these, and using the massless condition, the branch cuts
disappear in the overall amplitude. Including universal geometric
factor of sphere $8 \pi g_c^{-2} \alpha'^{-1} $, we have the
overall factor 
\begin{equation}
 2 (4 \pi)^{-L+3} g_{10}^{L-2} e^{(L-2)\Phi}\alpha'^{\frac12(L-4)},
\end{equation}
up to contributions from picture changing and/or
oscillator excitations (\ref{vertexnorm}). For the four point 
coupling, we have volume factor suppression from the compactification.  

As the simple example, we can extract the information for
three point correlation function. By $SL(2,\C)$
symmetry, we can fix three arbitrary given points completely, thus
the correlation function contains no information. To obtain it, we
need four point correlation function with twists
\begin{equation}
 \langle \sigma_{k}(0,0),
\sigma_{N-k}(x,\bar x), \sigma_{l}(1,1),
\sigma_{N-l}(\infty, \infty) \rangle ,
\end{equation}
where we consider only one two-torus. From (\ref{wmatrix}), we have
\begin{equation} \begin{split}
 \det F(x,\bar x)
 = & \textstyle  F_1^1 F_2^2 - F_1^2 F_2^1  \\
 = & \textstyle  B(1-\frac kN,\frac lN)
 F(1-\frac{k}{N},\frac{l}{N};1+\frac{l}{N}-\frac{k}{N};1-z)
 F(\frac{k}{N},1-\frac{l}{N};1;\bar z)  \\
 \textstyle &- \text{  ($k \leftrightarrow l, z \leftrightarrow \bar z$
 exchanged) }.
\end{split} \end{equation}
Let us assume $k+l<N$. In the limit $z,\bar z \to \infty$, using the
relations (\ref{zinftyhyper1}) and (\ref{zinftyhyper2}) in Appendix,
the most dominant part has the coefficient
\begin{equation}
\det F \to \pi { \Gamma^2(1-\frac{k+l}{N}) \over  \Gamma^2(1-\frac{k}{N})
 \Gamma^2(1-\frac{l}{N}) } \left[ { e^{-ik\pi} \over \sin{\pi l
 \over N}} + {e^{il\pi} \over \sin{\pi k \over N}} \right]^{-1} ,
\end{equation}
where we used the relation $\Gamma(a) \Gamma(1-a) = \pi / \sin a
\pi$.
Thus, taking the entire compact dimension, the Yukawa coupling is
obtained as
\begin{equation} \label{Yukawa}
Y^{k,l}_{f_a,f_b,f_c} = g_{\rm YM,4D} \prod_{j=1}^{d/2} (2 A_{\Lambda,j})^{1/2}
{\Gamma(1-\frac{k_j}{N}) \Gamma(1-\frac{l_j}{N}) \over
 \Gamma(1-\frac{k_j+l_j}{N})} \left[ { e^{-ik_j\pi} \over \sin{\pi l_j
 \over N}} + {e^{il_j\pi} \over \sin{\pi k_j \over N}}
 \right]^{1/2} \sum_{v} \exp(-S_{\cl})
\end{equation}
with the classical action given in (\ref{3ptclaction}).
For the case $k_j+l_j>N$, we can
obtain the corresponding amplitude by replacing $k_j,l_j$ by $N-k_j,N-l_j$.
It is  notable that, in the quantum amplitude, there is no
contribution from the geometry, since it is canceled by the same
dependence in the four dimensional gauge coupling arising from the
dimensional reduction.

We can obtain similar expressions for the higher order couplings,
expressed in terms of multivalued hypergeometric functions.
One can be convinced that the asymptotic form of generalized
hypergeometric function in the above limit is the ratios of gamma
functions \cite{FL} with arguments of ${\cal O}(0.1)$. Thus we expect 
that the quantum parts of the
higher order couplings are roughly of order one. Thus
we see that the size of higher order coupling is dominated by
the classical part.

\section{Conclusions}

We have calculated couplings of arbitrary order, among untwisted and
twisted fields of heterotic string on orbifolds, using conformal field
theory. In the low energy limit, they correspond to the matter
superpotential. They are given by the zero external momentum limit of
radially ordered correlation functions.
The specification of orbifold and the shift vector determines the possible
couplings. This provides us with lessons for constructing low-energy
effective field theory, in particular for vacuum configurations of
compactified string theory and realistic quark and lepton masses.

The higher order couplings are complicated due to two things. The
first is the technical difficulties dealing with arbitrary number of
twist fields and even some excited twist fields from the picture
changing. In the calculation of higher order coupling, 
the latter has just
an effect of changing the normalization, not changing 
the transformation property and the branch cut
structure. The other difficulty arises from linearly
combined states, which appear in higher twisted sectors of non-prime
order orbifolds. They make the interpretation of the classical and
the quantum somewhat tricky.

The selection rules (mainly studied in Ref. \cite{CKNRV}) from the
location of fields and the 
$R$-symmetry can be possible origins of discrete quantum numbers 
in effective field theory. They
also provides the understanding on discrete flavor symmetries.
Because of the restrictive form of selection rules, in particular in the
heterotic string models, only limited number of couplings are
possible, since there are limited number of orbifolds and fixed
points. This will be classified elsewhere \cite{CK}.

The classical part is an instanton amplitude of world-sheet nature.
Its size is exponentially suppressed by the effective area swept
by the twisted strings to form untwisted strings. The latter is
energetically not allowed, metastable intermediate state. This
generalizes the naive area rule even if the twisted strings do not
form a polygon. Decomposing the locations of twisted fields, we
can obtain handy rule for calculating the size. For couplings
involving linearly combined states, the only contribution comes from
the terms satisfying the space group selection rule. The others vanish
individually. 

We have also calculated the quantum amplitude with the complete
normalization, up to the K\"ahler normalization. In this there is no
contribution from the geometric distribution of the fields, since
the contributions from the normalization and that from the dimensional
reduction cancel.

Besides the $\alpha'^{1/2}$ suppressions from the closed string
couplings and oscillator normalizations, the coefficient is given by
ratios of products of gamma function, whose argument is of $O( 0.1)$,
thus we expect a factor of $O(1)$ from the quantum amplitude.
The dominant amplitude is classical one, which is
exponentially suppressed. Thus it is easy to generate hierarchy of
Yukawa couplings. However, from the top-down approach, it is not
easy to locate the desired fields at the desired positions.

\subsection*{Acknowledgements}
We are grateful to Steve Abel, Massimo Bianchi, Jihn
E. Kim, Hans-Peter Nilles, Saul Ramos-Sanchez, Michael Ratz, 
Robert Richter and Patrick K.~S. Vaudrevange 
for discussion. K.~S.~C. thanks to Yukawa
Institute for hospitality, where part of the work is done.
K.~S.~C. is supported in part by the European Union 6th framework program
MRTN-CT-2004-503069 ``Quest for unification",
MRTN-CT-2004-005104 ``ForcesUniverse", MRTN-CT-2006-035863
``UniverseNet'' and SFB-Transregio 33 ``The Dark Univerese"
by Deutsche Forschungsgemeinschaft (DFG).
T.~K.\/ is supported in part by the Grand-in-Aid for
Scientific Research \#1754025 and
the Grant-in-Aid for the 21st Century
COE ``The Center for Diversity and Universality in Physics'' from the
Ministry of Education, Culture, Sports, Science and Technology of Japan.

\appendix

\section{Useful formulae}

The four point amplitude is described by the standard hypergeometric
functions. 
The following relations 
\begin{align} \label{gamma1minusz}
  F(a,b;c;z) =&
   {\Gamma(c)\Gamma(c-a-b)\over
  \Gamma(c-a)\Gamma(c-b)}F(a,b;a+b+1-c;1-z) \nonumber \\
  &+ {\Gamma(c)\Gamma(a+b-c)\over\Gamma(a)\Gamma(b)}
   (1-z)^{c-a-b}F(c-a,c-b;1+c-a-b;1-z) ,     \\
  \label{gammasamec}
F(a,b;c;z)  =&(1-z)^{c-a-b}F(c-a,c-b;c;z) ,
\end{align}
are useful.

The classical action contains the following functions,
\begin{align} \label{f0}
   F_0(1-x) & = -(-1)^{-k_4/N} \textstyle B(\frac{k_4}{N},\frac{k_1}{N})
 F(\frac{k_4}{N},1-\frac{k_2}{N},\frac{k_1}{N}+\frac{K_4}{N};x), 
\nonumber \\
  F_1(x)  & \textstyle = (-1)^{(k_2+k_3)/N} x^{-1+(k_1+k_2)/N}
  B(\frac{k_1}{N},\frac{k_2}{N})
  F(\frac{k_1}{N},1-\frac{k_3}{N};\frac{k_1}{N}+\frac{k_2}{N};x), 
\nonumber \\
  F_2(1-x) & = \textstyle (-1)^{-1+k_3/N} (1-x)^{-1+(k_2+k_3)/N}
  B(\frac{k_2}{N},\frac{k_3}{N})
  F(\frac{k_3}{N},1-\frac{k_1}{N};\frac{k_2}{N}+\frac{k_3}{N};1-x), 
\nonumber \\
%\end{align}
%\begin{align}
  F_3(x) & = \textstyle B(\frac{k_3}{N},\frac{k_4}{N})
 F(\frac{k_4}{N},1-\frac{k_2}{N},\frac{k_3}{N}+\frac{k_4}{N};x),\\
  F_1'(\bar x) & = F_1(k_i \to N-k_i;\bar x), \nonumber \\
  F_2'(1-\bar x) &  = F_2(k_i \to N-k_i;1-\bar x),  \nonumber
%\label{f2prime}
\end{align}
where $B(a,b)=\Gamma(a)\Gamma(b)/\Gamma(a+b)$ and it is 
the Euler beta function.
For $F_0$ and $F_3$ we have used above relations
(\ref{gamma1minusz}),(\ref{gammasamec}), as well as (\ref{sumk}).
For $\sum k_i = 2 \pi N$, the above functions become as 
\begin{align} \label{f2prime}
   F_0(1-x) & =  {\sin(\pi k_3/N) \over \sin(\pi k_4/N)} e^{i \pi
 (k_3+k_4)/N} \left[
  {\sin(\pi(k_2+k_3)/N) \over \sin(\pi k_1/N)} e^{ik_2 \pi/N}
 F_1(x) + F_2(1-x) \right],  \nonumber \\
%\end{align}
%\begin{align}
   F_3(x) & =  {\sin(\pi k_1/N) \over \sin(\pi k_4/N)} e^{i \pi
 (k_2+k_3)/N} \left[
 F_1(x) + {\sin(\pi(k_1+k_2)/N) \over \sin(\pi k_1/N)} e^{-ik_2 \pi/N}
 F_2(1-x) \right], \nonumber \\
  F_1'(\bar x)  & =  (-1)^{(k_2+k_3)/N}
 \bar  x^{1-(k_1+k_2)/N} (1-\bar x)^{1-(k_2+k_3)/N} \nonumber \\
    & \quad \times \textstyle B(1-\frac{k_1}{N},1-\frac{k_2}{N})
  B(\frac{k_3}{N},\frac{k_4}{N})^{-1} F_3 (\bar x), \\
  F_2'(1-\bar x)  & =  (-1)^{1-(k_3+k_4)/N} (1-\bar x)^{1-(k_2+k_3)/N}
  \bar x^{1-(k_1+k_2)/N} \nonumber \\
  & \quad \times \textstyle B(1-\frac{k_2}{N},1-\frac{k_3}{N})
  B(\frac{k_1}{N},\frac{k_4}{N})^{-1} F_0 (1-\bar x). \nonumber 
\end{align}

For factorization of the amplitude, we need the asymptotic behaviors of
hypergeometric functions in the $z \to \infty$
limit
\begin{align}
  F(a,b;c;z) & \simeq e^{\pi i a} {\Gamma(c) \Gamma(b-a) \over \Gamma(b)
  \Gamma(c-a)} z^{-a} + e^{\pi i b} {\Gamma(c)\Gamma(a-b) \over
  \Gamma(a)\Gamma(c-b) } z^{-b} , \label{zinftyhyper1}  \\
 F(a,b;c;1-z) & \simeq {\Gamma(c) \Gamma(b-a) \over \Gamma(b)
  \Gamma(c-a)} z^{-a} + {\Gamma(c)\Gamma(a-b) \over
  \Gamma(a)\Gamma(c-b) } z^{-b}. \label{zinftyhyper2}
\end{align}

For the higher order amplitude than four, we need a multivariable
generalization of hypergeometric function, called Lauricella D
function \cite{Ext}. It is defined as
\begin{equation}
 F_D(a,b_1,\dots,b_r;c;x_1,\dots,x_r) \equiv \sum_{m_1=0}^\infty \dots
 \sum_{m_r=0}^\infty {(a)_{m_1+\dots+m_r} (b_1)_{m_1} \dots
 (b_r)_{m_r} \over (c)_{m_1+\dots+m_r} m_1! \dots m_r!} x_1^{m_1}
 \dots x_r^{m_r},
\end{equation}
with $|x_i|<1$ for all $i$. Here $(a)_n$ is the
Pochhammer symbol meaning
\begin{equation} \begin{split}
  (a)_n &= {\Gamma(a+n) \over \Gamma(a)}  = a(a+1) \dots (a+n-1), \quad n \ge
  0, \\
  (a)_n &= {(-1)^{-n} \over (1-a)_{-n}}, \quad n<0.
\end{split} \end{equation}
We can express the integration as \cite{Abel:2003yx}
\begin{equation}
  \begin{split}
    F_i^1 =& e^{i \pi k_i/N}(x_i - x_{i+1})^{-1+(k_i+k_{i+1})/N}
    \prod_{j=1,j \ne i,i+1}^{N-1} (x_i-x_j)^{-1+k_j/N}
    B(k_i/N,k_{i+1}/N) \\
&\times
F_D^{(N-3)}(k_i/N,1-k_1/N,\dots,1-k_{L-1}/N;k_i/N+k_{k_i+1}/N;
 \tilde x_{i,1},\dots \tilde x_{i,L-1}),\\
 F_0^1 =& e^{-i \pi(k_N/N+1)} B(k_N/N,k_1/N) \\
&\times
F_D^{(N-3)}(k_N/N,1-k_2/N,\dots,1-k_{N-2}/N;k_1/N+k_N/N;1-x_2,
\dots,1-x_{N-2}),
  \end{split}
\end{equation}
where $\tilde x_{ij}={x_i - x_{i+1} \over x_i - x_j}$ is the
conformally invariant cross ratio.

\end{document}